\documentclass[10pt, conference]{IEEEtran}
\IEEEoverridecommandlockouts
\usepackage{cite}
\usepackage{amsmath,amssymb,amsfonts}
\usepackage{algorithmic}
\usepackage{graphicx}
\usepackage{textcomp}
\usepackage{xcolor}
\usepackage{fancyvrb}
\DeclareGraphicsExtensions{.png,.pdf}
\fvset{%
fontsize=\small,
numbers=left
}

\def\BibTeX{{\rm B\kern-.05em{\sc i\kern-.025em b}\kern-.08em
    T\kern-.1667em\lower.7ex\hbox{E}\kern-.125emX}}

\begin{document}

\title{Analyzing Impact of Dependency Injection on Software Maintainability}

\author{\IEEEauthorblockN{Peter Sun and Dae-Kyoo Kim}
\IEEEauthorblockA{\textit{Computer Science and Engineering} \\
\textit{Oakland University}\\
Rochester, MI, USA \\
\{pysun,kim2\}@oakland.edu}
}

\maketitle

\begin{abstract}
Dependency injection (DI) is generally known to improve maintainability by keeping application classes separate from the library. Particularly within the Java environment, there are many applications using the principles of DI with the aim to improve maintainability. There exists some work that provides an inference on the impact of DI on maintainability, but no conclusive evidence is provided. The fact that there are no publicly available tools for quantifying DI makes such an evidence more difficult to be produced. In this paper, we propose a novel metric, DCBO, to measure the proportion of DI in a project based on weighted couplings. We describe how DCBO can serve as a more meaningful metric in computing maintainability when DI is also considered. The metric is implemented in the CKJM-Analyzer, an extension of the CKJM tool that utilizes static code analysis to detect DI. We discuss the algorithmic approach behind the static analysis and prove the soundness of the tool using a set of open-source Java projects.
\end{abstract}


\section{Introduction}
Software development has grown increasingly dependent on external libraries that are built by other companies. While it provides convenience in development, it significantly increases the cost of software maintenance especially for the changes that involve the dependency with libraries. The use of external libraries also requires significant overheads such as extra code to be imported and compiled,
resulting in performance bottlenecks.

Component frameworks (e.g., Spring~\cite{Johnson04}) help mitigate development cost. A key feature of component frameworks for object-oriented programming (OOP) is dependency injection (DI). DI is a pattern of sending ("injecting") necessary fields ("dependencies") into an object, instead of requiring the object to initialize those fields itself.
Existing literature~\cite{Razina07, Yang08, Lee00, Crasso10} suggests
that the use of DI can help improve the maintainability of software systems. On the other hand, there are also warnings against
using DI due to possible negative effects~\cite{Roubtsov10, Laigner19}.

Razina and Janzen~\cite{Razina07} compared the maintainability of Spring projects and non-Spring projects where Spring projects are assumed to have DI. However, as they mentioned in their work, they found no obvious correlation between the use of DI and maintainability. While their initial hypothesis that Spring projects would have lower coupling than non-Spring projects was disproved, they did see a few projects in the Spring group that had a significant decrease in coupling correlating to an increase in DI. However, we think that their hypothesis was rejected because it is possible -- and more than likely -- that non-Spring projects also have DI. This is difficult to assess in their work because how DI was counted was not described, which is critical for determining whether the non-Spring projects had DI considered in the same way as Spring projects. For example, if Razina and Janzen only analyzed XML files for DI (as implied in their work), it is more likely that non-Spring projects can implement DI in non-XML configured methods. 

In this paper, we present a novel metric, Dependency Injection-Weighted Coupling Between Objects (DCBO) and CKJM-Analyzer~\cite{ckjm-analyzer}, a command-line interface (CLI) extending the original CKJM tool~\cite{CKJM} to analyze DI and compute metric means across all classes within Java projects.
DCBO weighs coupling between objects depending on whether it is soft-coupled (e.g., with DI) or hard-coupled (e.g., with
the \verb|new| keyword or with using an object generator that
requires parameter information from the user).
We developed CKJM-Analyzer to more easily iteratively analyze Java projects for CKJM metrics and incorporate algorithms to compute the proportion of DI in Java classes. We believe CKJM-Analyzer will be very useful for statistical analysis on Java projects with respect to CKJM metrics.
We validate the metric and tool with a set of open-source Java projects.

The remainder of the paper is organized as follows. Section~\ref{sec:relatedwork} gives an overview of the related work on the effect of DI in software systems, as well as work in measuring coupling weight. Section~\ref{sec:dependency-injection} presents a background on DI. Section~\ref{sec:approach} describes DCBO and its algorithmic approach implemented in the CKJM-Analyzer tool.
Section~\ref{sec:data-analysis} describes the data analysis results from CKJM-Analyzer on small sample projects and open-source projects. Section~\ref{sec:discussion} discusses the results with regards to the impact of DI in maintainability, the effect of DCBO on coupling analysis, limitations and potential future work. Section~\ref{sec:conclusion} concludes the paper with discussion on future research work.

\section{Related work}
\label{sec:relatedwork}
Razina and Janzen~\cite{Razina07} studied the impact of DI on maintainability. They count the number of DIs in a Spring project (i.e., a project built on the Spring framework which is built on DI ~\cite{Johnson04}) and divide the number by the sum of CBOs of the project for normalization. Maintainability is measured in terms of CBO, RFC, and LCOM using the CKJM tool~\cite{CKJM}. They formed two groups of projects -- Spring projects and non-Spring projects with an assumption that Spring projects use the DI pattern and non-Spring projects do not. They conducted ANOVA analysis on those groups to find out correlation between the use of DI and maintainability. However, they found no obvious correlation between the use of DI and maintainability. 

While their initial hypothesis that Spring projects would have lower coupling than non-Spring projects was disproved, they did see a few projects in the Spring group that had a significant decrease in coupling correlating to an increase in DI. We think that their hypothesis was rejected because it is possible - and more than likely - that non-Spring projects also have DI. This is difficult to assess in Razina and Janzen's paper because they did not describe how DI was counted, which is critical for determining whether the non-Spring projects had DI considered in the same way as Spring projects. For example, if Razina and Janzen only analyzed XML files for DI, it is more likely that non-Spring projects can implement DI in non-XML configured methods. They considered non-weighted CBO to evaluate the effect of DI. 
We think non-weighted CBO is insufficient to determine the effect of DI. 
Section~\ref{sec:approach} describes DCBO and why we believe it more accurately assesses the effect of DI on coupling and maintainability, and section~\ref{sec:data-analysis} details experimental results that show unweighted CBO remaining the same regardless of DI proportion in experimental projects.

The work by Crasso et al.~\cite{Crasso10} suggests that the use of DI can provide cleaner web-based interfaces. The model they developed demonstrates a significant improvement in the precision of interfaces for web applications with DI applied. An improvement was observed in web service queries, and the authors think it is due to DI code containing more ``meaningful terms'' or having higher accuracy in words/phrases that relate to search queries. Additionally, the authors observed a small performance hit and overhead increase when using DI, primarily because of the DI pattern requiring more memory in service adaptors. The authors argue that the slight increase in memory and overhead is minimal, especially because DI-based web services provide more accurate queries. The authors do not directly mention maintainability, but suggest that DI makes it easier to outsource Web Service development.

There exists also some work~\cite{Roubtsov10, Saidulu14} discussing a negative impact of DI. The work by Roubtsov et al.~\cite{Roubtsov10} presents bad smells caused by DI on modularity. They observed that a specialized form of DI using syntactic metadata has a higher probability of violating modularity rules, which leads to less cohesion. Similarly, the work by Laigner et al.~\cite{Laigner19} presents a catalog of DI anti-patterns (e.g., {\em framework coupling}, {\em intransigent injection}) which increase coupling (e.g., by DI annotations, unnecessary dependencies) while decreasing maintainability as opposite to ``good'' patterns (e.g., GoF patterns~\cite{Gamma95}). 

Saidulu~\cite{Saidulu14} proposed a metric coined Weighted Coupling Between Objects (WCBO). They rank the weight of class coupling with bipartite graphs and use the metric to assess fault proneness. Our work differs from Saidulu's approach in that we weight coupling via DI.

\section{Dependency Injection}
\label{sec:dependency-injection}
DI, which is a specific form of {\em Dependency Inversion Principle}~\cite{Yang08}, is a design pattern to improve the maintainability of software systems by reducing coupling through injecting dependencies in classes using an external injector which is a class object or file (e.g, an XML-based configuration file in the Spring framework).
As coupling is reduced. consequently the complexity of classes is also dminished. DI also makes it easier to pinpoint dependency-related errors as dependency injection is localized in one place (viz. the injector).

According to Yang et al~\cite{Yang08}, a dependency may be injected in four ways~\cite{Yang08} -- (i)  via constructor parameters, which is known as {\em Constructor No Default} (CND); (ii) via method parameters, which is known as {\em Method No Default} (MND); (iii) via constructor parameters or a default object (using the \verb|new| command), which is known as {\em Constructor With Default} (CWD); and  (iv) via method parameters or via a default object, which is known as {\em Method With Default} (MWD). Consider the code snippet below. In the code, the \verb|DogPenCND| class implements CND, the \verb|DogPenMND| class implements MND, the \verb|DogPenCWD| class implements CWD, and the \verb|DogPenMWD| class implements MWD.

\begin{Verbatim}
public class Dog {
    String name;
    Dog() {
        this.name = "Dog";
    }
}

public class DogPenCND {
    Dog dog;
    DogPen(Dog dog) {
        this.dog = dog;
    }
}

public class DogPenMND {
    Dog dog;
    void AddDog(Dog dog) {
        this.dog = dog;
    }
}

public class DogPenCWD {
    Dog dog;
    DogPen() {
        this.dog = new Dog();
    }
}

public class DogPenMWD {
    Dog dog;
    void AddDog() {
        this.dog = new Dog();
    }
}
\end{Verbatim}

Typically, the implementation of services is specified in the injector which is often used as a clue for the use of DI. When a change needs to be made in the service, it is done through the injector without changing the client. In this way, the client remains unchanged and thus, the code becomes more
flexible and reusable. 

Consider the code snippet below. The \verb|DogPenGeneratorCND| class acts as the injector, injecting the \verb|Dog| class into each \verb|DogPenCND| object. In this way, each different \verb|DogPenCND| object does not need to generate its own \verb|Dog| dependency. This is particularly helpful when the same dependency is injected in multiple different objects. The CND and MND
structures follow this benefit.

\begin{Verbatim}
public class DogPenGeneratorCND() {
    Dog dog = new Dog("Dog1");
    DogPenCND pen1 = new DogPenCND(dog);
    DogPenCND pen2 = new DogPenCND(dog);
    DogPenCND pen3 = new DogPenCND(dog);
}
\end{Verbatim}

In contrast, the CWD and MWD structures do not entirely follow the injector benefit. Consider the code snippet below. If developers use the "default" constructor or method available in the CWD/MWD structures, it requires the object itself to generate its own dependency object. 

\begin{Verbatim}
public class DogPenGeneratorCWD() {
    DogPenCWD pen1 = new DogPenCWD();
    DogPenCWD pen2 = new DogPenCWD();
    DogPenCWD pen3 = new DogPenCWD();
}
\end{Verbatim}

DI comes with a few technical hindrances. Firstly, it requires all the dependencies to be resolved before compilation if the compiler is not configured to recognize injected dependencies. That is, the compiler cannot recognize the presence of injected dependencies unless it is configured. Secondly, the frameworks built upon DI are often implemented with reflection or dynamic programming, which can hinder the use of IDE automation such as finding references, showing call hierarchies, and safe refactoring~\cite{Fowler04}.

\section{Analyzing the Impact of DI on Maintainability}
\label{sec:approach}
In this section, we describe the proposed approach for analyzing the impact of DI on software maintainability. The following are the research questions we seek to answer in this work. 

{\em Is CBO a sufficient metric to determine DI in a system?} We believe this is an important question to address because previous research has argued that DI should reduce coupling, and therefore reduce CBO~\cite{Razina07}. It should follow, therefore, that an inverse relationship between DI and CBO (increase in DI will decrease CBO) can be quantified. We use experimental analysis to determine whether that inverse relationship exists.

{\em Does weighted DI in a system yield greater maintainability?} This is the motivation behind our proposal for the new metric, DCBO. We show in Section~\ref{sec:data-analysis} that CBO values are consistent when Java projects are implemented with any of the four DI definitions proposed by Yang et al.~\cite{Yang08}, and we further show how DCBO - which weighs the definitions differently - can provide a more meaningful metric in determining the effect of DI on maintainability.

In particular, we hope to expand on the work done by 
Razina and Janzen~\cite{Razina07} who focused on Spring framework
projects to assess whether DI improves maintainability. We aim
to provide an open-source software solution in computing
DI within projects, and propose a new metric in accordance with
an updated DI definition.

\subsection{Dependency Injection Weighted-Coupling Between Objects}
Yang et al.~\cite{Yang08} acknowledged a reduction in 
reusability with respect to the CWD and MWD patterns because of the 
``default'' object, but did not believe that was significant enough to remove the CWD and MWD definitions~\cite{Yang08}.
We argue that only CND and MND conform to the DI pattern (cf. Section ~\ref{sec:dependency-injection}) and with respect to CWD and MWD, the ``default'' object incurs twice the coupling penalty
compared to injecting dependencies via parameters. In other words, DI via parameter injection (CND/MND~\cite{Yang08})
requires only one change where the dependency is generated
and injected into various classes, whereas DI via default object injection
(CWD/MWD~\cite{Yang08}) requires {\em N} number of changes,
where {\em N} is the number of classes that depend on that object.
We propose DCBO to quantify the effect of DI on coupling
in Java classes. We weigh coupling in two parts:

\begin{itemize}
	\item Half of the total weight of 2 is added if a coupled dependency exists (i.e., half of the total CBO metric). A total weight of 2 is added every time a coupling exists, because of the two-way nature of the CBO metric computation.
	\item Half of the total weight of 2 is added if the dependency the class is coupled to is not injected via DI (i.e., the class uses a default object).
\end{itemize}

Given that, we quantify DCBO using the following equation, where $DIP$ is the number of parameters that are injected via DI using the CND/MND~\cite{Yang08} pattern.

\begin{equation}\label{def:dcbo}
	DCBO = CBO - DIP
\end{equation}

CKJM-Analyzer was developed to address a need for an open-source solution that can quantify DI in Java projects. At a high level, it is a .NET
command line interface (CLI)~\cite{ckjm-analyzer} run on a Windows operating system (OS) that wraps the CKJM extended tool~\cite{Chidamber94} and provides additional functionality to compute DI proportion, project-specific metric means, and DCBO. 
In the CKJM-Analyzer tool, we compute DCBO for every class
using Formula~\ref{def:dcbo}, and finally return the mean
for every project in the final report.
We acknowledge that while the weights assigned in DCBO
should sufficiently cover DI implementations via parameter
injection, it does not account for whether that injected
object will enforce further coupling. For example,
the injected object could call an internal function that
requires multiple parameters, forcing the user to perform
changes wherever that function is called, even though that
object is injected via parameter injection. In that case,
DCBO would compute its coupling weight as 0.5, but
practically the weight should remain as 1 since the user
will need to perform multiple changes.

\subsection{Dependency Injection Detection Algorithm}
In this section, we discuss the specific algorithm implemented to detect DI. It identifies DI patterns analogous to CND and MND, but does not consider the default object injection as a DI pattern (viz. CWD and MWD). CKJM-Analyzer iteratively executes CKJM-DI on every Java class
within a project and performs additional computations. Below is pseudocode for the algorithmic implementation to detect DI. In the code,
{\em GetParams()} returns a list of all parameters the class uses in its constructors or methods. Then, {\em GetClassNames()} returns a list of all class names, essentially all of the available non-primitive dependencies in the project. The intersection of \verb|params| and \verb|classNames| is a list of all non-primitive parameters that the class has, which follows CND/MND DI pattern. Then, \verb|diPara| is multiplied by 2 when computing the proportion because coupling is a two-way relation (i.e., the object using the dependency and the dependency itself will each increment coupling by 1). Given that the computation returns a proportion (\verb|di|), we deem the lowest DI proportion to be 0 (0\%) and the highest DI proportion to be 1 (100\%). 

\begin{Verbatim}
double diPara = 0.0;
double cbo = 0.0;
foreach classFile in project:
    var params = GetParams();
    var classNames = GetClassNames();
    diPara += Intersect(params, classNames);
    cbo += GetCBO();
double di = diPara * 2 / cbo;
\end{Verbatim}

\subsection{Maintainability Algorithm}
Maintainability is defined as 
\emph{``the ease with which a software system or component can be modified to correct faults, improve performance or other attributes, or adapt to a changed environment''--IEEE.}
There have been many models proposed to quantify maintainability~\cite{Lincke08, Oman92, Wagey15},
but we propose a simplified model based on metrics analyzed by Razina and Janzen~\cite{Razina07}. We believe comparing normalized metrics across projects would more accurately reflect how maintainability changes when DI proportion changes.
We compute maintainability using the CBO, LCOM and RFC metrics. We then normalize CBO as NCBO using the module complexity formula proposed by Okike~\cite{Okike2008}, $CM = 1 - \frac{1}{1 + IS}$, where \verb|CM| is the module complexity (normalized CBO), and \verb|IS| is the coupling complexity (CBO). We also normalize RFC as NRFC using the module complexity formula. We normalize LCOM as NLCOM using the Best-Fit normalization approach~\cite{Okike2008}, where \verb|NLCOM = 1/LCOM| if LCOM is not 0, otherwise \verb|NLCOM = 0| if LCOM is 0. Given that, maintainability can be computed as MAI.

\begin{equation}\label{eq:mai}
 MAI = 1 - \frac{NCBO}{3} - \frac{NLCOM}{3} - \frac{NRFC}{3}   
\end{equation}

where the value of \verb|MAI|, \verb|NCBO|, \verb|NLCOM|, and \verb|NRFC| is in the range of [0, 1]. A \verb|MAI| score of 1 is the highest maintainability, and a \verb|MAI| score of 0 is the lowest maintainability. As CBO, LCOM, and RFC increase, the system becomes more complex and less maintainable. That is, a system with very high CBO, LCOM and RFC will result in a system that is extremely difficult to maintain. However, \verb|MAI| does not take into account dependency injection and thus, it cannot capture the impact of dependency injection on maintainability. So, we substitute NCBO with NDCBO (normalized DCBO) to consider dependency injection. We propose DMAI (Dependency Injection-Weighted Maintainability);

\begin{equation}\label{eq:dmai}
 DMAI = 1 - \frac{NDCBO}{3} - \frac{NLCOM}{3} - \frac{NRFC}{3}   
\end{equation}

CKJM-Analyzer parses the metric results from CKJM-DI, computes the metric means, and calculates maintainability with Formula~\ref{eq:mai}, as well as dependency injection-weighted maintainability with Formula~\ref{eq:dmai}.

\section{Data Analysis}
\label{sec:data-analysis}
In this section, we use CKJM-Analyzer to analyze artificially generated Java projects, as well as open-source projects.

\subsection{Experimental Data}
We first test CKJM-Analyzer and DCBO using artificially generated Java projects. In an effort to showcase CKJM-Analyzer use and DCBO, we created simple Java projects~\cite{ckjm-di-projects} and performed DI analysis using our tools (viz. CKJM-Analyzer and CKJM-DI)~\cite{ckjm-analyzer, ckjm-di}. Each project contains a \verb|Dog| class and ten \verb|DogPen| classes. The projects range from {\em di\_0} that has every \verb|DogPen| class initializing its own \verb|Dog| dependency class (following the CWD pattern~\cite{Yang08}), and {\em di\_100} that has every \verb|DogPen| class using an injected \verb|Dog| dependency class (following the CND pattern~\cite{Yang08}). A total of 11 projects were created, and CKJM-Analyzer was executed on the projects.

Table~\ref{tab:non-norm-exp-results} shows the results of non-normalized metrics from the application. Of the metrics included in the MAI metric, only RFC sees a small decrease as DI increases. Both CBO and LCOM remain unchanged regardless of the amount of DI in the system. CBO stays consistent because it does not weigh CND differently from CWD, and sees all patterns of DI proposed by Yang et al.~\cite{Yang08} as the same type of coupling. Contrarily, DCBO has a noticeable decrease as DI increases - increasing DI From 0\% to 100\% decreases DCBO coupling by 50\%. DCBO weighs CWD more heavily than CND, and therefore shows a reduction in coupling as DI increases in the system.

\begin{table}[h]
\centering
\begin{tabular}{|c|c|c|c|c|c|c|}
\hline
Project & DI & CBO & DCBO & LCOM & RFC & LOC \\
\hline\hline
di\_0   & 0.0                    & 1.82                    & 1.82                     & 0                        & 2.91                    & 108                     \\
\hline
di\_10  & 0.1                    & 1.82                    & 1.73                     & 0                        & 2.82                    & 106                     \\
\hline
di\_20  & 0.2                    & 1.82                    & 1.64                     & 0                        & 2.73                    & 104                     \\
\hline
di\_30  & 0.3                    & 1.82                    & 1.55                     & 0                        & 2.64                    & 102                     \\
\hline
di\_40  & 0.4                    & 1.82                    & 1.45                     & 0                        & 2.55                    & 100                     \\
\hline
di\_50  & 0.5                    & 1.82                    & 1.36                     & 0                        & 2.45                    & 98                      \\
\hline
di\_60  & 0.6                    & 1.82                    & 1.27                     & 0                        & 2.36                    & 96                      \\
\hline
di\_70  & 0.7                    & 1.82                    & 1.18                     & 0                        & 2.27                    & 94                      \\
\hline
di\_80  & 0.8                    & 1.82                    & 1.09                     & 0                        & 2.18                    & 92                      \\
\hline
di\_90  & 0.9                    & 1.82                    & 1.00                     & 0                        & 2.09                    & 90                      \\
\hline
di\_100 & 1.0                    & 1.82                    & 0.91                     & 0                        & 2.00                    & 88                     \\
\hline
\end{tabular}
\newline\caption{CKJM-Analyzer Experimental Analysis Results}
\label{tab:non-norm-exp-results}
\end{table}

\begin{figure}[!htb]
\centering
	\scalebox{0.44}{\includegraphics{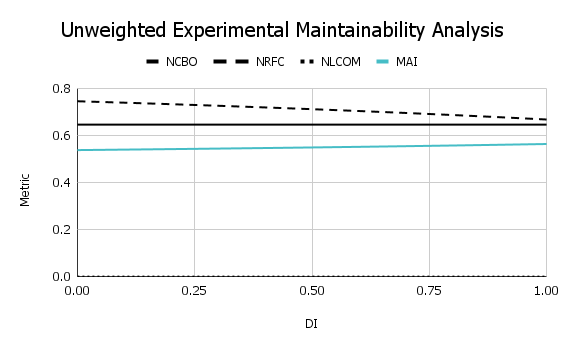}}
	\caption{Experimental Maintainability Analysis Using CBO}
	\label{fig:unweighted-exp-results}
\end{figure}

Table~\ref{tab:norm-unweighted-exp-results} shows the normalized metric and MAI results with unweighted CBO which are visualized in  Figure~\ref{fig:unweighted-exp-results}. The results show that given the small decrease in RFC as DI increases, there is a slight overall increase in MAI as DI increases.

\begin{table}[h]
\centering
\begin{tabular}{|c|c|c|c|c|}
\hline
DI & NCBO & NRFC & NLCOM & MAI \\
\hline\hline
0.0                    & 0.65                     & 0.74                     & 0                         & 0.54                    \\
\hline
0.1                    & 0.65                     & 0.74                     & 0                         & 0.54                    \\
\hline
0.2                    & 0.65                     & 0.73                     & 0                         & 0.54                    \\
\hline
0.3                    & 0.65                     & 0.73                     & 0                         & 0.54                    \\
\hline
0.4                    & 0.65                     & 0.72                     & 0                         & 0.55                    \\
\hline
0.5                    & 0.65                     & 0.71                     & 0                         & 0.55                    \\
\hline
0.6                    & 0.65                     & 0.70                     & 0                         & 0.55                    \\
\hline
0.7                    & 0.65                     & 0.69                     & 0                         & 0.55                    \\
\hline
0.8                    & 0.65                     & 0.69                     & 0                         & 0.56                    \\
\hline
0.9                    & 0.65                     & 0.68                     & 0                         & 0.56                    \\
\hline
1.0                    & 0.65                     & 0.67                     & 0                         & 0.56                   \\
\hline
\end{tabular}
\newline\caption{Normalized Unweighted CKJM-Analyzer Experimental Results}
\label{tab:norm-unweighted-exp-results}
\end{table}

\begin{figure}[!htb]
\centering
	\scalebox{0.44}{\includegraphics{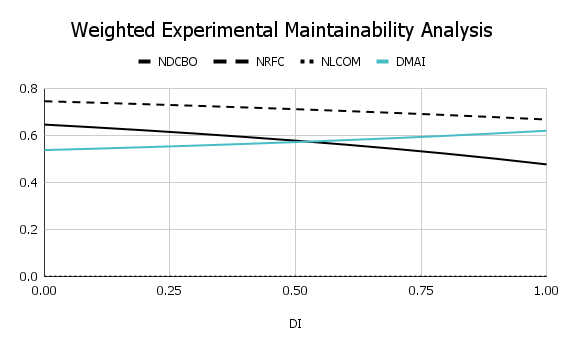}}
	\caption{Experimental Maintainability Analysis Using DCBO}
	\label{fig:weighted-exp-results}
\end{figure}

Table~\ref{tab:norm-weighted-exp-results} shows the normalized metric and MAI results with DCBO, and Figure~\ref{fig:weighted-exp-results} visualizes the results. The results show that NDCBO decreases as DI increases, which increases MAI more than the unweighted CBO results.

\begin{table}[h]
\centering
\begin{tabular}{|c|c|c|c|c|}
\hline
DI & NDCBO & NRFC & NLCOM & DMAI \\
\hline\hline
0.0                    & 0.64                      & 0.74                     & 0                         & 0.54                     \\
\hline
0.1                    & 0.63                      & 0.73                     & 0                         & 0.54                     \\
\hline
0.2                    & 0.62                      & 0.73                     & 0                         & 0.55                     \\
\hline
0.3                    & 0.60                      & 0.72                     & 0                         & 0.56                     \\
\hline
0.4                    & 0.59                      & 0.71                     & 0                         & 0.56                     \\
\hline
0.5                    & 0.57                      & 0.71                     & 0                         & 0.57                     \\
\hline
0.6                    & 0.56                      & 0.70                     & 0                         & 0.58                     \\
\hline
0.7                    & 0.54                      & 0.69                     & 0                         & 0.59                     \\
\hline
0.8                    & 0.52                      & 0.68                     & 0                         & 0.60                     \\
\hline
0.9                    & 0.50                      & 0.67                     & 0                         & 0.61                     \\
\hline
1.0                    & 0.47                      & 0.66                     & 0                         & 0.62                    \\
\hline
\end{tabular}
\newline\caption{Normalized Weighted CKJM-Analyzer Experimental Results}
\label{tab:norm-weighted-exp-results}
\end{table}

Finally, we perform Friedman statistical analysis~\cite{Garcia2008} by splitting the data set in half -- projects with DI $<0.5$ are labeled``No DI'', and projects with DI $>0.5$ are labeled ``DI'' -- and determining whether there is a significant difference with respect to MAI or DMAI. The output of the statistical tool described in Derrac et al.~\cite{Derrac2008} is displayed in Table~\ref{tab:exp-stats}.

According to the tool, when $\alpha=0.05$, Bergmann's procedure rejects ``No DI'' DMAI vs. ``DI'' DMAI, indicating significance in maintainability with respect to DI using DCBO. However, Bergmann's procedure does not reject ``No DI'' MAI vs. ``DI'' MAI, indicating no significance in maintainability with respect to DI using unweighted CBO.

\begin{table}[!htp]
\centering
\caption{Holm / Shaffer Table for $\alpha=0.05$}
\begin{tabular}{cccc}
algorithms&$p$\\
\hline
``No DI'' DMAI vs. ``DI'' DMAI&0.0143\\
``No DI'' MAI vs. ``DI'' MAI&0.0500\\
\hline
\end{tabular}
\newline\caption{Experimental Friedman Statistical Analysis}
\label{tab:exp-stats}
\end{table}

\subsection{Open-Source Data}
We test CKJM-Analyzer using open-source projects compiled from Yang et al.'s work~\cite{Yang08} and Tempero et al.'s work~\cite{Tempero2010}. We found the projects in online repositories as a jar file or a folder of class files.
There are several motivations for analysis. First, we want to see whether the experimental observations (CBO remains the same regardless of DI, while DCBO decreases while DI increases) are reproducible in real-world projects.
Second, we also want to assess the practicality of the proposed maintainability metric. 

Table~\ref{tab:ckjm-analyzer-os-results} shows the non-normalized metric results from the open-source analysis. The range of DI proportion present across projects is much smaller than the experimental results (0.1-0.5 instead of 0.0-1.0). 

\begin{table}[h]
\centering
\begin{tabular}{|c|c|c|c|c|c|c|}
\hline
Project & DI & CBO & DCBO & LCOM & RFC & LOC \\
\hline\hline
argouml       & 0.12                   & 9.89                    & 9.30                     & 36.49                    & 23.12                   & 262584                  \\
\hline
poi           & 0.17                   & 7.65                    & 6.99                     & 91.11                    & 26.55                   & 124773                  \\
\hline
ireport       & 0.20                   & 4.00                    & 3.59                     & 41.64                    & 15.35                   & 21378                   \\
\hline
wro4j         & 0.23                   & 7.00                    & 6.21                     & 23.27                    & 15.93                   & 29059                   \\
\hline
fitjava       & 0.25                   & 3.83                    & 3.34                     & 22.54                    & 16.49                   & 5178                    \\
\hline
jhotdraw      & 0.26                   & 8.70                    & 7.55                     & 77.32                    & 19.97                   & 66843                   \\
\hline
ant           & 0.26                   & 8.52                    & 7.40                     & 50.50                    & 25.94                   & 54268                   \\
\hline
hsqldb        & 0.28                   & 10.05                   & 8.66                     & 223.99                   & 37.85                   & 145646                  \\
\hline
advancedGwt   & 0.30                   & 8.38                    & 7.13                     & 122.34                   & 23.21                   & 33199                   \\
\hline
ganttproject  & 0.30                   & 6.66                    & 5.66                     & 133.69                   & 26.32                   & 66120                   \\
\hline
jGap          & 0.31                   & 8.10                    & 6.82                     & 77.25                    & 21.68                   & 970716                  \\
\hline
jchempaint    & 0.33                   & 8.63                    & 7.23                     & 144.77                   & 26.60                   & 880275                  \\
\hline
pdfBox        & 0.33                   & 11.10                   & 9.27                     & 68.27                    & 25.31                   & 158833                  \\
\hline
picocontainer & 0.34                   & 7.76                    & 6.44                     & 16.92                    & 14.93                   & 8931                    \\
\hline
mars          & 0.36                   & 7.33                    & 6.02                     & 28.50                    & 17.84                   & 108126                  \\
\hline
tomcat        & 0.36                   & 5.20                    & 4.26                     & 35.08                    & 19.29                   & 54343                   \\
\hline
jfreechart    & 0.36                   & 10.03                   & 8.20                     & 200.44                   & 33.22                   & 160004                  \\
\hline
jHotDraw      & 0.37                   & 5.89                    & 4.80                     & 98.71                    & 25.55                   & 370465                  \\
\hline
log4j         & 0.38                   & 6.04                    & 4.88                     & 44.27                    & 19.91                   & 36336                   \\
\hline
colt          & 0.48                   & 6.75                    & 5.13                     & 53.22                    & 18.73                   & 101730                  \\
\hline
xerces        & 0.50                   & 5.36                    & 4.02                     & 96.51                    & 20.36                   & 196673                 \\
\hline
\end{tabular}
\newline\caption{CKJM-Analyzer Open-Source Analysis Results}
\label{tab:ckjm-analyzer-os-results}
\end{table}

\begin{figure}[!htb]
\centering
	\scalebox{0.40}{\includegraphics{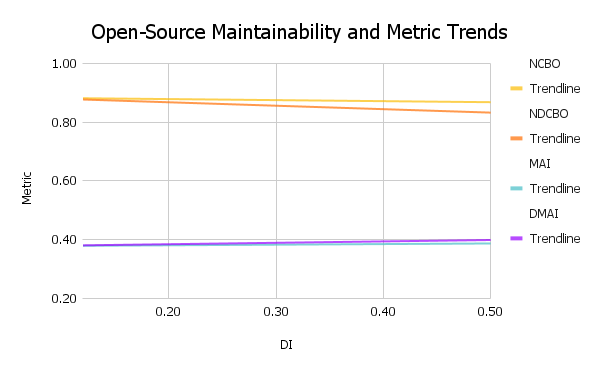}}
	\caption{Open-Source Maintainability and Coupling Analysis Trendlines}
	\label{fig:os-results}
\end{figure}

Table~\ref{tab:os-results-norm} shows the normalized metric results. In the results, small increases in MAI and DMAI as DI increases are observed. Figure~\ref{fig:os-results} visualizes the normalized coupling and maintainability results as trendlines. Similar to the experimental results, there is a greater decrease in NDCBO compared to NCBO as DI increases, and a greater increase in DMAI compared to MAI as DI increases.

\begin{table}[h]
\centering
\begin{tabular}{|c|c|c|c|c|c|c|}
\hline
DI & NCBO & NLCOM & NRFC & MAI & NDCBO & DMAI \\
\hline\hline
0.12                   & 0.91                     & 0.03                      & 0.96                     & 0.37                    & 0.90                      & 0.37                     \\
\hline
0.17                   & 0.88                     & 0.01                      & 0.96                     & 0.38                    & 0.87                      & 0.38                     \\
\hline
0.20                   & 0.80                     & 0.02                      & 0.94                     & 0.41                    & 0.78                      & 0.42                     \\
\hline
0.23                   & 0.88                     & 0.04                      & 0.94                     & 0.38                    & 0.86                      & 0.38                     \\
\hline
0.25                   & 0.79                     & 0.04                      & 0.94                     & 0.41                    & 0.77                      & 0.41                     \\
\hline
0.26                   & 0.90                     & 0.01                      & 0.95                     & 0.38                    & 0.88                      & 0.38                     \\
\hline
0.26                   & 0.89                     & 0.02                      & 0.96                     & 0.37                    & 0.88                      & 0.38                     \\
\hline
0.28                   & 0.91                     & 0.00                      & 0.97                     & 0.37                    & 0.90                      & 0.37                     \\
\hline
0.30                   & 0.89                     & 0.01                      & 0.96                     & 0.38                    & 0.88                      & 0.39                     \\
\hline
0.30                   & 0.87                     & 0.01                      & 0.96                     & 0.39                    & 0.85                      & 0.39                     \\
\hline
0.31                   & 0.89                     & 0.01                      & 0.96                     & 0.38                    & 0.87                      & 0.39                     \\
\hline
0.33                   & 0.90                     & 0.01                      & 0.96                     & 0.38                    & 0.88                      & 0.38                     \\
\hline
0.33                   & 0.92                     & 0.01                      & 0.96                     & 0.37                    & 0.90                      & 0.37                     \\
\hline
0.34                   & 0.89                     & 0.06                      & 0.94                     & 0.37                    & 0.87                      & 0.38                     \\
\hline
0.36                   & 0.88                     & 0.04                      & 0.95                     & 0.38                    & 0.86                      & 0.39                     \\
\hline
0.36                   & 0.84                     & 0.03                      & 0.95                     & 0.39                    & 0.81                      & 0.40                     \\
\hline
0.36                   & 0.91                     & 0.00                      & 0.97                     & 0.37                    & 0.89                      & 0.38                     \\
\hline
0.37                   & 0.85                     & 0.01                      & 0.96                     & 0.39                    & 0.83                      & 0.40                     \\
\hline
0.38                   & 0.86                     & 0.02                      & 0.95                     & 0.39                    & 0.83                      & 0.40                     \\
\hline
0.48                   & 0.87                     & 0.02                      & 0.95                     & 0.39                    & 0.84                      & 0.40                     \\
\hline
0.50                   & 0.84                     & 0.01                      & 0.95                     & 0.40                    & 0.80                      & 0.41                    \\
\hline
\end{tabular}
\newline\caption{Normalized CKJM-Analyzer Open-Source Analysis Results}
\label{tab:os-results-norm}
\end{table}

We perform Friedman statistical analysis~\cite{Garcia2008} by splitting the data set in half. Given the nature of the data set, the cutoff point for the open-source projects differs from the cutoff point for the experimental projects -- projects $\le0.3$ are labeled ``No DI'', and projects $>0.3$ are labeled ``DI''. The analysis shows no significance for either MAI or DMAI, but a much lower $p$ value is observed when maintainability is analyzed using DCBO as compared to CBO.

\begin{table}[!htp]
\centering
\caption{Holm / Shaffer Table for $\alpha=0.05$}
\begin{tabular}{cccccc}
algorithms&$p$\\
\hline
NDI-DMAI vs. DI-DMAI&0.2253\\
NDI-MAI vs. DI-MAI&0.9999\\
\hline
\end{tabular}
\end{table}

\section{Discussion}
\label{sec:discussion}
In this section, we discuss the implications of the experimental and open-source project results, as well as future steps.

\paragraph*{Experimental results prove the soundness of CKJM-Analyzer and DCBO} According to Table~\ref{tab:non-norm-exp-results}, only RFC decreased as DI increased, while CBO and LCOM remained the same. Our argument that CBO does not weigh DI's impact on the system is shown given the coupling value did not change. More specifically, as we are now drawing a distinction between CND/MND and CWD/MWD patterns~\cite{Yang08}, the experimental results show that CBO views a dependency added via parameter injection the same as one added via \verb|new| initialization (the ``default'' object). In other words, CBO views coupling the same, regardless of whether class implements the CND/MND or the CWD/MWD pattern.

The Friedman analysis on the experimental data in Table~\ref{tab:exp-stats} shows a significant change in DMAI when comparing ``No DI'' projects with ``DI'' projects, but no significant change in MAI when comparing ``No DI'' projects with ``DI'' projects. This indicates that, when using DCBO as the coupling metric instead of CBO, a significant increase in maintainability is observed in systems with higher DI. Under the specific conditions where the primary difference in projects is the amount of DI they contain, DCBO would significantly impact maintainability, while CBO would not significantly impact maintainability.

\paragraph*{Open-source project results do not reflect experimental results} When splitting the data set in half to compare ``No DI'' against "DI" projects, no significant increase in maintainability was found using either CBO or DCBO. We believe there are a few reasons for this. Firstly, the range of DI proportion analyzed in open-source projects (0.1-0.5) was much lower than the range of DI proportion analyzed in experimental projects (0.0-1.0). Secondly, we did not consider XML-based DI, which may contribute to an increase in DI and DCBO. Lastly, we did not define a set cutoff for when to deem a project as implementing DI (i.e. how much DI should be present in the system to determine it uses DI). Given that the threshold was different from the experimental and open-source analyses, having a lower cutoff (0.3 for the open-source results compared to 0.5 for the experimental results) could reduce the impact on maintainability between ``No DI'' and ``DI'' projects.

\paragraph*{Different normalization techniques and metrics should be considered when computing maintainability} The normalization techniques employed on CBO, LCOM and RFC worked best on CBO because CBO values were closest to 1. LCOM had consistently high values, which normalized very close to 0, while RFC normalized very close to 1. Only CBO and DCBO saw substantial differences in their normalized values. In particular, LCOM has been viewed as a difficult metric to compare~\cite{Okike2008} because it is not normalized and the values have a high variance (as shown in the open-source results in Section~\ref{sec:data-analysis}). Future research should consider other techniques to provide more meaningful normalization, or limit comparisons to similar projects so that metrics can be better compared (similar to Razina and Janzen's work~\cite{Razina07}). 

\paragraph*{CKJM-Analyzer and DCBO may be useful in building more maintainable software} This paper covers only a small section of how DI can impact the maintainability of a project. In particular, we argued that DI via parameter injection instead of \verb|new| object ("default" object) construction improves maintainability by reducing the number of files that need to be changed should the injected class' constructor parameters change. Given the observation that DI does not impact CBO, CKJM-Analyzer and DCBO explicitly incorporate DI as a factor in weighing coupling differently and, therefore, impacting maintainability. One future work of CKJM-Analyzer would be determining which portions of the software are using the CWD/MWD patterns and replacing them with the CND/MND patterns~\cite{Yang08}. Another work would be making CKJM-Analyzer cross-platform as right now it is only usable on the Windows OS.

\section{Conclusion}
\label{sec:conclusion}
We have presented CKJM-Analyzer, a command line tool to analyze DI and maintainability, as well as DCBO, a weighted coupling metric that explicitly considers parameter-injected patterns of DI as a more maintainable framework. The algorithm used to detect DI is defined and the formulae to compute maintainability are discussed. The experimental results show that DI significantly improves maintainability when DCBO is considered, but does not significantly improve maintainability when CBO is considered. More research is needed to determine whether open-source projects show that same significance. We argue that DI may contribute to more maintainable software by reducing the developer "effort" when a dependency change is required.

\bibliographystyle{IEEEtran}
\bibliography{main}

\end{document}